\documentclass[twocolumn,prb,showpacs]{revtex4}
\usepackage{fancyhdr}
\usepackage{amsbsy}
\usepackage{amsmath}
\usepackage{subfigure}
\usepackage{epsfig}
\usepackage{amssymb}
\usepackage{units} 
\usepackage{pifont}
\usepackage{textcomp}
\usepackage{gensymb}
\usepackage{pifont}
\usepackage{enumerate}
\usepackage{color}
\begin{document}

\title{
Gap opening and large spin-orbit splitting in MX$_2$ (M=Mo,W X=S,Se,Te) from\\
the interplay between crystal field and hybridizations: insights from {\it ab-initio} theory
}

\author{Carmine Autieri$^1$, Adrien Bouhon$^1$ and Biplab Sanyal$^1$}

\affiliation{$^1$Department of Physics and Astronomy, Uppsala University, Box-516, 75120 Uppsala, Sweden,}


\date{\today}
\begin{abstract}
By means of first-principles density functional calculations, we study the maximally localized Wannier functions for the 2D transition metal dichalcogenides MX$_2$ (M=Mo,W X=S,Se,Te). We found a M$^{+4}$-like ionic charge and a single occupied $d$-band.
The center of the $d$-like maximally localized Wannier function associated with this band is distributed among three M sites.
Part of the energy gap is opened by the crystal field splitting induced by the X$^{-2}$-like atoms.
We extract the hopping parameters for the Wannier functions and provide a perspective on tight-binding model.
From the analysis of the tight binding model, we have found an inversion of the
band character between the $\Gamma$ and the $K$ points of the Brillouin zone due to the M-M hybridization.
The consequence of this inversion is the closure of the gap.
 The M-X hybridization is the only one that tends to open the gap at every k-point, the change in the M-X and M-M hybridization is the main responsible for the difference in the gap between the different dichalcogenide materials.
 The inversion of the bands gives rise to different spin-orbit splitting at $\Gamma$ and $K$ point in the valence band.
The different character of the gap at $\Gamma$ and $K$ point offers the chance to manipulate the semiconductive properties of these compounds.
For a bilayer system, the hybridizations between the out of plane orbitals and the hybridizations between the in plane orbitals split the valence band respectively at the $\Gamma$ and K point. The splitting in the valence band is opened also without spin-orbit coupling and comes from the M-M and X-X hybridization between the two monolayers.
\end{abstract}

\maketitle

\section{Introduction}

The synthesis of graphene has boosted
the research in atomically thin two-dimensional (2D) materials\cite{graphene}. The ability to manipulate
single atomic layers and reassemble them to form heterostructures opens novel routes for
 applications. In this field, 2D semiconducting dichalcogenides MX$_2$ (M=Mo,W X=S,Se,Te)
are promising compounds since they can
be easily exfoliated and present a suitable gap for electronic devices.\cite{Novoselov05,Olle,Mak10}
The dichalcogenides  systems
has recently gained attention for combining an electron mobility comparable to that of
graphene devices with a direct energy gap in monolayer and indirect gap in multilayers.
Another interesting feature is that the electronic properties and the gap size are
highly sensitive to external pressure, strain and temperature.\cite{properties1,properties2,properties3,properties4,Transition}
In addition, the lack of lattice inversion symmetry together with spin-orbit coupling (SOC) leads to
coupled spin and valley physics in monolayers of MX$_2$, making it possible to control spin and valley in these materials.

\begin{figure}[b!]
\centering
\includegraphics[width=8.7cm,height=8.3cm, angle=0]{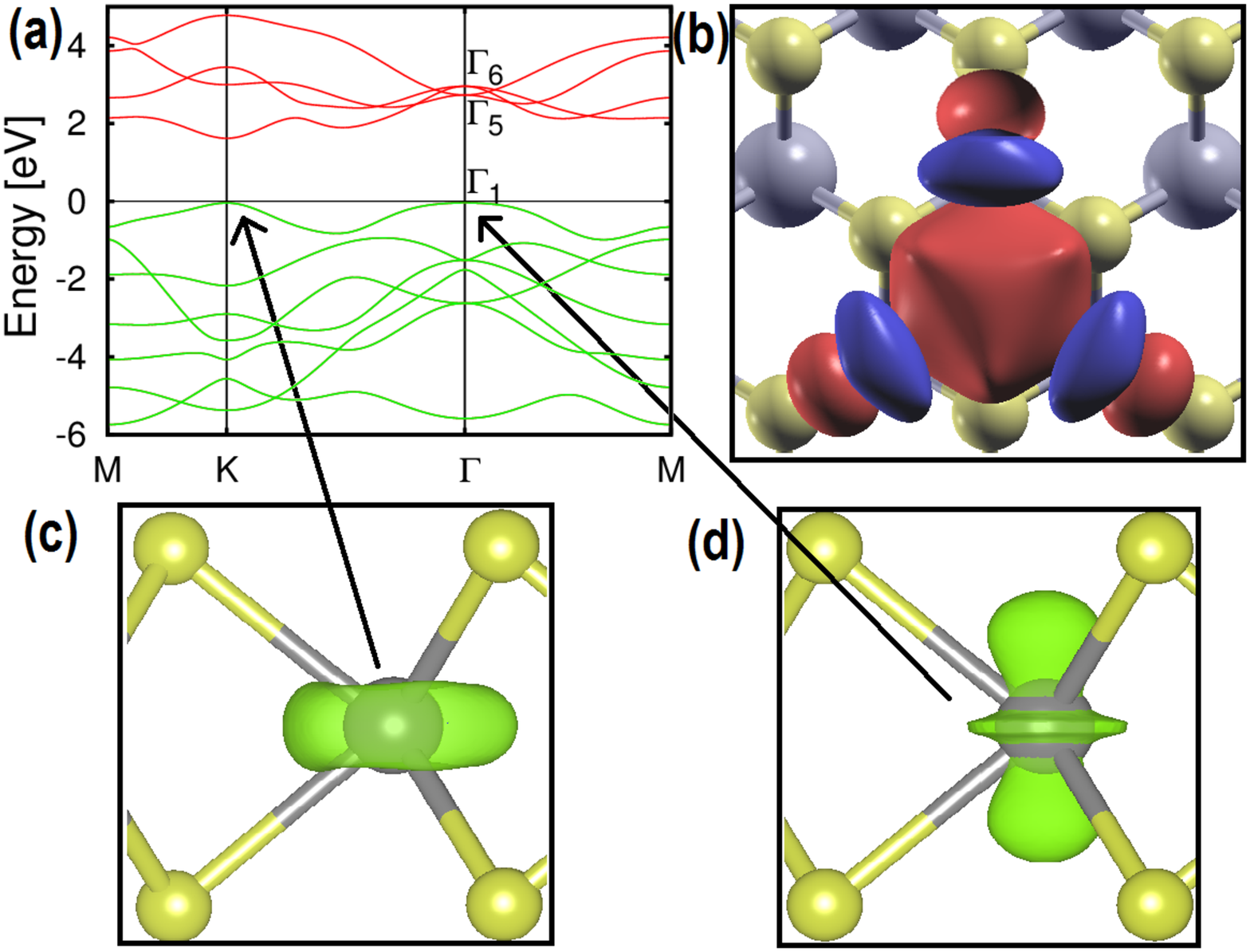}
\caption{Panel a). DFT band structure (red line) and interpolated band structure of the 7 highest occupied
bands (green line) obtained using the MLWF for MoS$_2$. The Fermi level is set to zero. Panel b). Top view of the MoS$_2$ single layer.
Mo and S atoms are shown as grey and yellow balls respectively.
We show the $d$-like Wannier function where red and blue contours are for isosurfaces of identical absolute values but opposite signs.
Panel c). Charge density of the valence band at the K point. Panel d). Charge density of the valence band at the $\Gamma$ point.
}
\label{MLWF}
\end{figure}
While for graphenic materials the theoretical work
has been based on tight binding-like approaches, this is not so easy for the MX$_2$ because of the larger number of bands.
Numerous tight binding (TB) models for MX$_2$ materials have been proposed in the last years. Most of them were derived using the Slater-Koster parametrization~\cite{Cappelluti,Zahid,Rostami,Ridolfi,Shanavas}. The difficulty of this approach to reproduce accurately the DFT band structure with a limited set of orbitals leads to the necessity of a trustful TB model. Even the crystal field is not accurately reproduced~\cite{Cappelluti,Zahid,Ridolfi}. A TB model based on the lattice symmetry just for the reflection-symmetric M-$d$ orbitals was derived~\cite{3bands} leading to a good match with the DFT energy bands. However including the X-$p$ orbitals is crucial in order to explain the electronic properties of MX$_2$ dichalcogenides. Without a good tight-binding model that accounts both for the M-$d$ and X-$p$ orbitals it is hazardous to deduce effective $\boldsymbol{k} \cdot\boldsymbol{p}$ models from a mere analogy with monolayer graphene~\cite{Xiao}. It is also not clear whether an analogy with bilayer graphene is relevant~\cite{Kormanyos}. In this work, we derive a TB model from group theory for the M-$d$ and X-$p$ orbitals and fix its free parameters using the matrix elements obtained from the MLWFs. This sets the basis for a future derivation of a full and accurate tight binding model and the corresponding effective $\boldsymbol{k} \cdot\boldsymbol{p}$ models.

From the TB parameters obtained from density functional theory,
we will study how the interplay between the crystal field and the hybridizations opens the gap
and produces different spin-orbit (SO) splitting in different regions of the Brillouin zone.
We will show how the lack of inversion symmetry produces the large SO splitting in the monolayer
and the coexistence of the spin orbit and hybridization contribution to the splitting in the bilayer.

\section{Computational details}

We have performed first-principles density functional theory (DFT)
calculations by using the VASP \cite{VASP2} package based
on plane wave basis set and projector augmented wave method.\cite{VASP}
A plane-wave  energy cut-off of 400~eV has been used.
For the treatment of exchange-correlation, Perdew-Burke-Ernzerhof \cite{PBE}
generalized gradient approximation has been considered.
After obtaining the Bloch wave functions in density functional theory, the maximally localized Wannier functions\cite{Marzari97,Souza01} (MLWF)
are constructed using the WANNIER90 code.\cite{Mostofi08}
A TB model based on group theory is derived.

We want to mention that the mismatch between our TB model and the DFT bands comes from two sources.
The first is the truncation of the range of hopping parameters.
The second source comes from a residual mixing between M and X orbitals inherent to Wannier90 (this is the consequence of the fact that the algorithm doesn't use any symmetry constraints) while in our model we work with disentangled orbitals. This issue will be addressed elsewhere. However since the M-X mixing of the MLWFs is quantitatively small, the incorporation of the matrix elements obtained by Wannier90 into our TB model leads to a good qualitative picture.

\section{Maximally localized Wannier functions for 7 occupied bands}

We define the last occupied band as the valence band (V) and the first unoccupied band as the conduction band (C).
We calculate the MLWFs for the last 7 occupied bands of the dichalcogenide materials MX$_2$
and find that the general features are similar for all of them.
Our results show the presence of 6 equivalent $p$-like MLWFs centred on the X atoms and 1
$d$-like in the basal plane. Therefore the other 4 $d$-bands are above the Fermi level.
Basically from ionic charge point of view we have a M$^{+4}$-like and two X$^{-2}$-like atoms.
This ionic picture it is more difficult to be understood by the local density of states.\cite{BandStrucutre}
 The presence of the ionic bond was already found in some compounds\cite{properties1}.
 The X$^{-2}$-like atoms are less electronegative than oxygen,
therefore the chemical bonds are slightly more covalent respect to oxides.
The 6 equivalent $p$-like MLWFs have the lobe of the orbital along the M-X direction.\cite{SUPMAT1} The only $d$-like MLWF shown in Fig. \ref{MLWF}-b) is more unusual.
It is a trivalent bond centered among three M atoms and it extends in the basal plane with C$_3$ symmetry.
One of the components of the gap in these systems is the crystal field splitting induced by X$^{-2}$-like atoms. The $d$-like MLWF avoids the $p$-like out of plane orbitals and it is lower in energy than the other $d$-orbitals.
Going down the group in the periodic table, the electronegativity of the X element decreases.
However the effective electronic charge on the
X atom is mostly unmodified and the crystal field (CF) persists.
Now, we want to investigate in more detail the $d$-character of the V band
and we try do disentangle the V band from the $p$-bands.

\section{Maximally localized Wannier functions for the valence band}

We repeat the MLWF procedure just for the V band. A tight binding (TB) model for the V band is presented in the Supplementary materials.\cite{SUPMAT2}
The V band presents an almost pure 3z$^2$-r$^2$ character at $\Gamma$ (see Fig. \ref{MLWF}-d) and a linear combination of x$^2$-y$^2$ and xy character at $K$ (see Fig. \ref{MLWF}-c).
On the contrary the M point has a mixed character with non negligible $p$-orbital contribution.
The corresponding MLWF lies mainly in the basal plane and it is basically the same of Fig. \ref{MLWF}-b).

\begin{figure}[!]
\centering
\includegraphics[height=6.8cm,width=8.6cm]{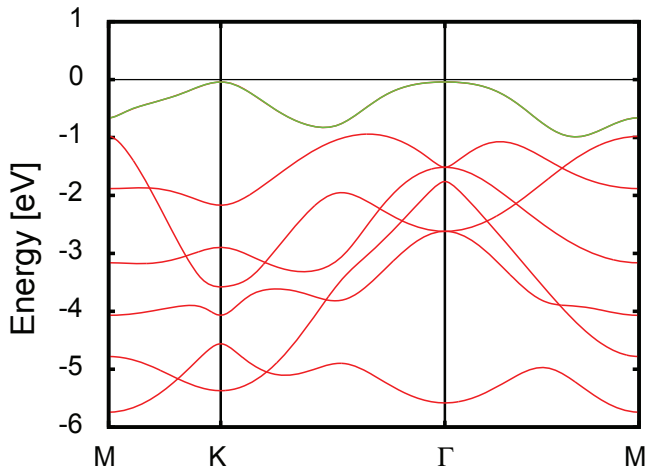}
\caption{DFT band structure of the 7 highest occupied bands (red) and the interpolated V
band (green) obtained using the MLWF for MoS$_2$. The Fermi level is set to zero.
}
\label{MLWFMoS2}
\end{figure}

\begin{figure}[!]
\centering
\includegraphics[height=6.8cm,width=8.6cm]{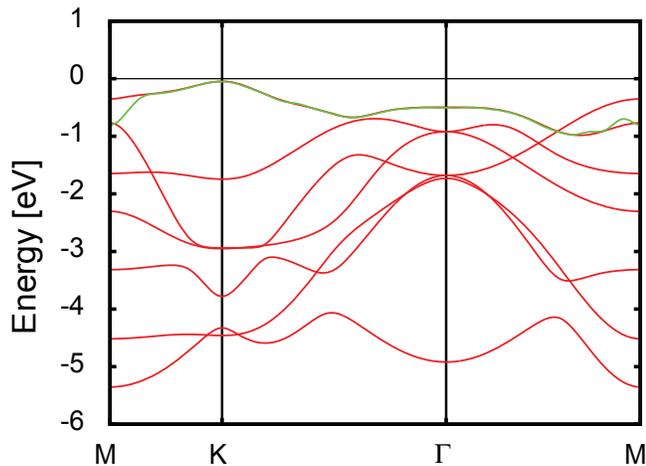}
\caption{DFT band structure of the 7 highest occupied bands (red) and the interpolated V
band (green) obtained using the MLWF for MoTe$_2$. The Fermi level is set to zero.
}
\label{MLWFMoTe2}
\end{figure}

We have found a perfect disentanglement between the V band and the other bands for the S/Se compounds while it is not possible to disentangle it for the Te-compounds as we can see in Fig. \ref{MLWFMoS2} and \ref{MLWFMoTe2}.
This is due to the strong $p$-character at the M point for the Te systems.
Due to the reduced electronegativity the Te electrons hybridize so strongly
with the V band that is not possible to disentangle them.
As a result, it is possible to derive an effective 3 bands model for the S/Se compounds as already proposed\cite{3bands} but not for the Te compounds. The different behaviour
of the Te compounds was already addressed regarding the optical properties and the formation energies.\cite{Som}

\section{Tight binding model}

A TB model based on group theory is derived for the basal plane reflection-symmetric orbitals only, including M-M,
M-X and X-X nearest-neighbor and next-nearest-neighbor hopping terms without the inclusion of SOC.
The lattice of monolayer MX$_2$ can be obtained from the combination of the translations of the 2D hexagonal lattice together with the rotations and reflections of the point group $D_{3h}$=$\{E,2 C_3,3 C_2',\sigma_h,2 S_3,3 \sigma_v \}$. It has no inversion center, hence the parity is not a good quantum number. Instead the reflection symmetry with respect to the basal plane, $\sigma_h$, plays an important role. We restrict ourselves to the $d$ electrons of M and the $p$ electrons of X atoms.
Since there is in the unit cell the bottom (B) and the top (T) X atoms, we introduce the bonding ($+$) and antibonding ($-$) basis $\psi_{X}^{\pm}\sim \psi_{X,T}\pm\psi_{X,B}$. At the $\Gamma$-point, the little group is $D_{3h}$ and the electronic orbitals split under the CF as (we use the notation of Ref.~\onlinecite{Koster})
\begin{equation}\label{SYM}
\begin{array}{cc}
	\begin{array}{rcl}
		\Gamma^{d_{x^2-y^2},d_{xy}} &=& \Gamma_6(\mathrm{M}) \\
		\Gamma^{d_{3z^2-r^2}} &=& \Gamma_1(\mathrm{M}) \\
		\Gamma^{p_{x},p_{y},+} &=& \Gamma_6(\mathrm{X}) \\
		\Gamma^{p_z,-} &=&\Gamma_1(\mathrm{X})  \\
	\end{array} 
	&
	\begin{array}{rcl}
		\Gamma^{d_{xz},d_{yz}} &=& \Gamma_5(\mathrm{M}) \\
		\Gamma^{p_{x},p_{y},-} &=& \Gamma_5(\mathrm{X}) \\
		\Gamma^{p_z,+} &=& \Gamma_4(\mathrm{X}) \\
	\end{array}
\end{array}
\end{equation}
with, in the left column, the states that are symmetric ($sym$) under $\sigma_h$, and, in the right column, the states that are antisymmetric ($asym$) under $\sigma_h$. Since translational symmetry here only involves the basal plane, the coupling between $sym$- and $asym$-states is not allowed. On the contrary, the coupling among $sym$- or $asym$-states is allowed. Below we focus on the $sym$-bands since they are the most relevant in order to understand the formation of the gap.

\begin{figure}[!]
\centering
\includegraphics[height=5.5cm,width=8.5cm]{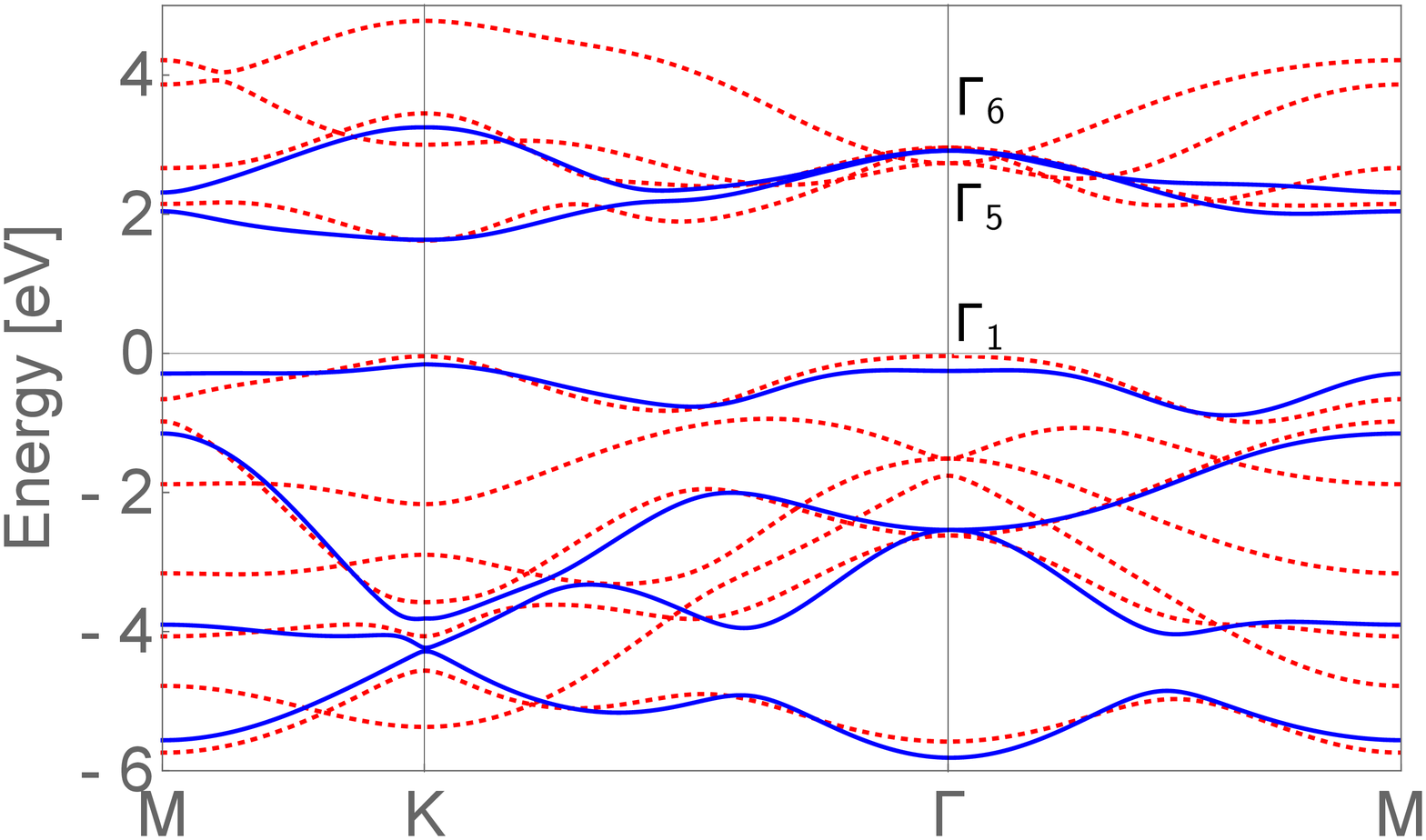}
\includegraphics[height=5.7cm,width=8.5cm]{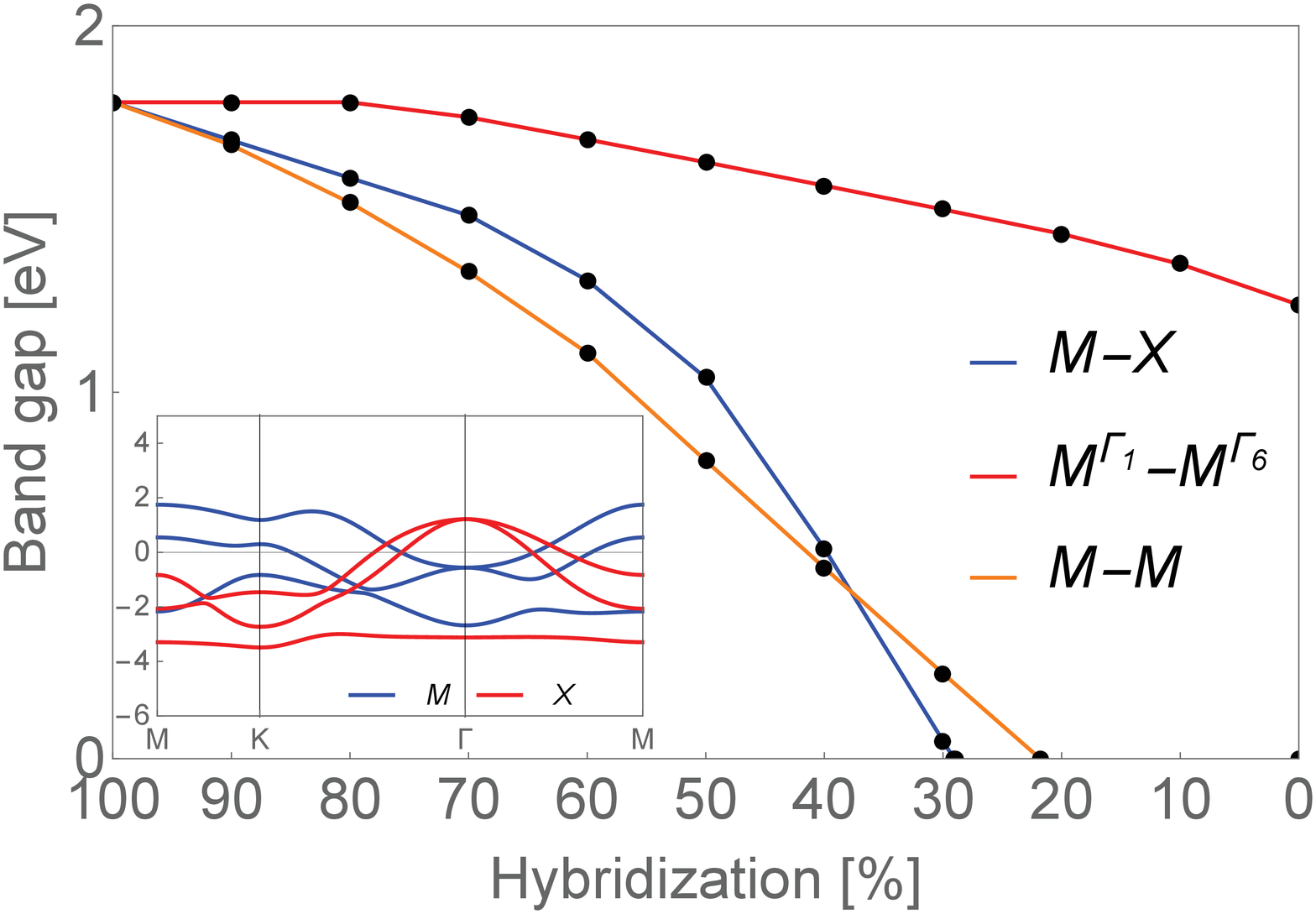}
\caption{DFT band structure (dotted red) and $sym$-bands obtained using the TB model (blue) for MoS$_2$ (top panel).
Band gap as a function of the M-X, M$^{\Gamma_1}$-M$^{\Gamma_6}$ and M-M hybridization (bottom panel).
In the inset we have the $sym$-bands obtained using the TB model with M-X hyvridization set to zero.
}
\label{TB}
\end{figure}

The comparison between the TB model and the DFT calculation for MoS$_2$ is reported in Fig. \ref{TB} (top panel).
We note the presence of twofold degeneracies of the band structure at the $\Gamma$-point which correspond to the two-dimensional representations introduced in Eq.~(\ref{SYM}), $\Gamma_6$ from the $sym$-bands and $\Gamma_5$ from the  $asym$-bands. As we go away from the $\Gamma$-point, the little group has a lower symmetry ($C_{3h}$ at the $K$-point, and $C_{2v}$ at the $M$-point) such that the two-dimensional representations split into two distinct one-dimensional representations.

We extract the hopping parameters for the model with the X-p and M-d orbitals. We
 analyze what happens at the $\Gamma$ point for the d-orbitals $sym$-bands.
The orbital for $\alpha$=1,2,3 are respectively x$^2$-y$^2$, xy and 3z$^2$-r$^2$.
Because of the orthogonality condition, the d-subspace of the hamiltonian is diagonal.\cite{SUPMAT3}
It is determined by the on-site energy and the M$^\alpha$-M$^\alpha$ hybridization
that are given in Table \ref{TabCF}.
\begin{table}[!ht]
\caption{The on-site energies and
sum of hopping integrals between the M atoms for the nearest-neighbor (NN), for the next-nearest-neighbor (NNN) as the selected MLWFs of MoS$_2$
respectively in the first, second and third line.
The on-site energy of the  3z$^2$-r$^2$-like WF is set to zero.
In the fourth and fifth lines, we report the diagonal matrix elements at $\Gamma$ and $K$ point up to the NN\cite{SUPMAT3}.
Hopping integrals $t_{0,i}^{\alpha,\alpha}$
from the site 0 with orbital $\alpha$ to neighboring site i with orbital $\alpha$.
$\varepsilon_0^\alpha$ is the on-site energy for the orbital $\alpha$. The unit is meV.}
\begin{center}
\begin{tabular}{|c|c|c|c|c|c|c|c|c|c|}
\hline
  & \multicolumn{3}{|c|}{MoS$_2$} & \multicolumn{3}{|c|}{MoTe$_2$} \\
\hline
                                                                          & x$^2$-y$^2$ & xy  & 3z$^2$-r$^2$ &  x$^2$-y$^2$ & xy  & 3z$^2$-r$^2$ \\
\hline
$\varepsilon_0^{\alpha}$                                                  &     499     &  536   &    0    &    484     &   506  &      0    \\
\hline
$\sum_{i=NN}t_{0,i}^{\alpha,\alpha}$                                     &    -722     & -764   & -1970   &   -385     &  -411  &  -1282    \\
\hline
$\sum_{i=NNN}t_{0,i}^{\alpha,\alpha}$                                     &    308     &  324   &  38     &    194     &   204  &    18      \\
\hline
$\varepsilon_0^{\alpha}$+$\sum_{i=NN}t_{0,i}^{\alpha,\alpha}$            &    -223     & -228   & -1970   &     99     &    95  &  -1282    \\
\hline
$\varepsilon_0^{\alpha}$-$\frac{1}{2}\sum_{i=NN}t_{0,i}^{\alpha,\alpha}$  &    860     &  918   &   985   &    677     &   712  &    641    \\
\hline
\end{tabular}
\end{center}
\label{TabCF}
\end{table}
From the on-site energy $\varepsilon_0^{\alpha}$ in Table \ref{TabCF}, we observe that the 3z$^2$-r$^2$ band is 0.5 eV lower in energy than the other d-bands.
This is in contradiction to what was found in a previous tight-binding model that used the Slater-Koster parametrization\cite{Cappelluti}. There the free parameters were obtained through a variational fitting of the M-$d$ and X-$p$ DFT bands.
We limit our considerations to the nearest neighbors and next nearest neighbors hybridization. The next nearest neighbors hybridization for the 3z$^2$-r$^2$ band is small as
for the p$_z$ orbitals in graphene.
The presence of the in plane orbitals xy and x$^2$-y$^2$ with their strong NNN hybridization
makes impossible to have an accurate TB model using just the NN hybridization as in the graphene case.

We can decompose the contributions to the direct gap at the $\Gamma$ point as the following.
From Table \ref{TabCF} we get: $\Delta^{CF}$=0.5~eV, $\Delta^{{M^\alpha}-{M^\alpha}}_{NN}$=1.2~eV and $\Delta^{{M^\alpha}-{M^\alpha}}_{NNN}$=0.3~eV.
Since the direct gap  for the MoS$_2$ at the $\Gamma$ point is 3.0~eV,
the exceeding part can be attributed to the M-X hybridization after the diagonalization of the hamiltonian giving $\Delta^{M-X}$=1.0~eV. At the $\Gamma$ point the CF, the M-M and M-X hybridization play in the same direction to open the gap with the M-M and M-X hybridization that give the largest contributions.
While the CF is barely affected if we move to MoTe$_2$, the M-M and M-X hybridization present a generalized homogenous reduction
of the hopping parameters.
Because the electronic structure, the crystal structure and the ionic charge do not change drastically, the reduction
of the hybridization can be mostly attributed to the increase in the lattice parameter. The reduction of the hybridization is responsible for the reduction of the gap in MoTe$_2$.

When we move from the $\Gamma$ point to the $K$ point the M$^\alpha$-M$^\alpha$ hybridization
changes sign as we can see from the formulas in fourth and fifth line of Table \ref{TabCF}.
From the CF in the first line and from the CF+M$^\alpha$-M$^\alpha$ at the $\Gamma$ point in the fourth line of Table \ref{TabCF},
the 3z$^2$-r$^2$ band is lower in energy.
Instead we can see from the last line of the Table \ref{TabCF} that the
energy of the 3z$^2$-r$^2$ band is 985 meV for MoS$_2$  at the K point and becomes
 higher than the in plane orbitals.
The M$^\alpha$-M$^\alpha$ hybridization produces a crossing of the bands
with the inversion of the band characters respect to the CF
and the closure of the gap.
The gap is reopened from the M-X hybridization. In order to understand if the M-X hybridization can sustain alone the gap we set the coupling between M and X to zero, as it is shown in the inset of the bottom panel of Fig.~\ref{TB}. We see a strong band inversion between the M-$d$ bands (blue) and the X-$p$ bands (red). As the M-X hybridization sets in, a gap opens as a consequence of the avoided crossings.
To check the importance of the M-X and M-M hybridization we show in Fig.~\ref{TB} (bottom panel) the effect on the gap of the different terms of the TB Hamiltonian. A reduction of the M-X and M-M hybridization shrinks the band gap as
we can observe respectively from the blue and orange line in Fig.~\ref{TB}.
This is confirmed by the strong dependence of the gap on the lattice parameter\cite{Transition}.
When the M-X hybridization is 29\% or the M-M hybridization is 22\% of the original value the gap is closed.
We conclude that both hybridizations are necessary in order to open the gap.
We can observe how the hybridization between the orbital belonging to the $\Gamma_1$ and $\Gamma_6$ representation is less relevant in these compounds. Looking at the Fig.~\ref{TB}, the gap does not close when this hybridization is zero.

After the inversion of the d-bands, the CF plays opposite to the opening of the gap.
At the $K$ point the strong M$^{xy}$-M$^{x^2-y^2}$ hybridization is effective,
this will create a bonding-antibonding scenario favoring the opening of the gap.
If we now decompose the energy gap for MoS$_2$ we get:
$\Delta^{CF}$=-0.5~eV, $\Delta^{{M^\alpha}-{M^\alpha}}_{NN}$=0.6~eV, $\Delta^{{M^\alpha}-{M^\alpha}}_{NNN}$=0.0~eV and $\Delta^{xy,x^2-y^2}_{NN}$=1.0~eV. Since the direct gap at the $K$ point is 1.7 eV, the exceeding part can be attributed to the M-X hybridization after the diagonalization of the hamiltonian giving $\Delta^{M-X}$=0.6~eV.


The bonding-antibonding scenario created by the M$^{xy}$-M$^{x^2-y^2}$ hybridization
is also the key point to understand the large SO splitting in the V band.
Indeed at the K point the hamiltonian of the xy/x$^2$-y$^2$ subsector for up and spin channel
can be approximated as\cite{SUPMAT3}
\[ \hat{H}(K) \simeq \left( \begin{array}{cc}
(t+\lambda)\sigma_2 & 0 \\
0 & (t-\lambda)\sigma_2   \end{array} \right)\]
where $\sigma_2$ is the Pauli matrix and $\lambda$ is the spin-orbit coupling constant.
The SO splitting is $\Delta E$=2$\lambda$.
This kind of hamiltonian is possible to construct due to the absence of the inversion symmetry.
From our M$^{xy}$-M$^{x^2-y^2}$ hybridization we get $t$=1008 meV for MoS$_2$,
this large value suppresses the other hybridizations and produces as eigenstates the
spherical harmonics (see Fig. \ref{MLWF}-c) with consequent SO splitting 2$\lambda$.
The M-X hybridization produces also a mixing of the M and X states in the V band at the K-point.
Following the mixing formula of the SO splitting\cite{SOC} the mixing favors the SO if $2\lambda_M<\lambda_X$.
In Table \ref{TabSOC} we report the spin orbit constants for the MX$_2$ compounds calculated from the band structure interpolation.
Using our estimations for the SOC, the M-X hybridization favors the SO for MoSe$_2$ and MoTe$_2$,
while is against it for the other compounds.

\begin{table}[!ht]
\caption{Spin orbit constants of the M ($\lambda_M$) and X element ($\lambda_X$) for the selected MX$_2$ compounds.
The unit is meV.
}
\begin{center}
\begin{tabular}{|c|c|c|c|c|c|c|}
\hline
                 & MoS$_2$ & MoSe$_2$  & MoTe$_2$ & WS$_2$ & WSe$_2$ & WTe$_2$ \\
\hline
 $\lambda_M$   &   85  & 86   & 81   & 276    & 279    &  256   \\
 $\lambda_X$   &   47  & 239  & 506  &  27   &  213   &   442  \\
\hline
\end{tabular}
\end{center}
\label{TabSOC}
\end{table}

The results of Table \ref{TabSOC} are comparable with previous results in the literature\cite{SOC}.
$\lambda_M$ slightly increases when we go from MS$_2$ to MSe$_2$ while decreases for MTe$_2$.
$\lambda_X$ decreases when we go from MoX$_2$ to WX$_2$.


We summarize in Table \ref{TabGap} how the interplay between CF, M$^{\alpha}$-M$^{\alpha}$, M$^{xy}$-M$^{x^2-y^2}$ and M-X hybridization opens the gap
in different regions of the Brillouin zone and produces the large SO splitting at the K-point in the V band.
The different properties of the gap in in different regions of the Brillouin zone offer the chance to manipulate the semiconductive properties of this compound.
\begin{table}[!ht]
\caption{The CF and the hybridizations can be favorable (F), against (A) or
ineffective (I) to generate the reported properties in the MX$_2$ monolayer.}
\begin{center}
\renewcommand{\arraystretch}{1.7}
\begin{tabular}{|c|c|c|c|c|}
\hline
                                     & CF &  M$^{\alpha}$-M$^{\alpha}$  & M$^{xy}$-M$^{x^2-y^2}$ &  M-X  \\
\hline
Gap at the $\Gamma$                  &    F    &  F   &  I   &   F     \\
\hline
Gap along the $\Gamma$-K line           &    F    &  A   &  A   &   F     \\
\hline
Gap at the $K$                         &    A    &  F   &  F   &   F     \\
\hline
Assists SO splitting                   &    I    &  I   &  F   &  A/F     \\
\hline
\end{tabular}
\end{center}
\label{TabGap}
\end{table}

\section{Bilayer MX$_2$}

Now, we analyze the case of the bilayer systems.
The distance between the two MX$_2$ layers is large due to the Coulomb repulsion between the X$^{-2}$-like atoms of different layers.
In the first approximation, we can just consider the effects of the additional monolayers acting on the M-orbitals
present at the Fermi level.
We use the prime($'$) for the quantity induced from the other layers.
There are four contributions: the CF$'$, the M-X$'$, the M-M$'$ and the X-X$'$ hybridization.
The first two are found to be always negligible.
The main effect on the M-orbitals is the X-X$'$ hybridization via M-X producing
a bonding-antibonding scenario in the M-orbitals.\cite{Cappelluti}
The M-M$'$ hybridization is one order of magnitude smaller but plays in the same direction as X-X$'$.
\begin{figure}[b!]
\centering
\includegraphics[width=8.6cm,height=5.0cm, angle=0]{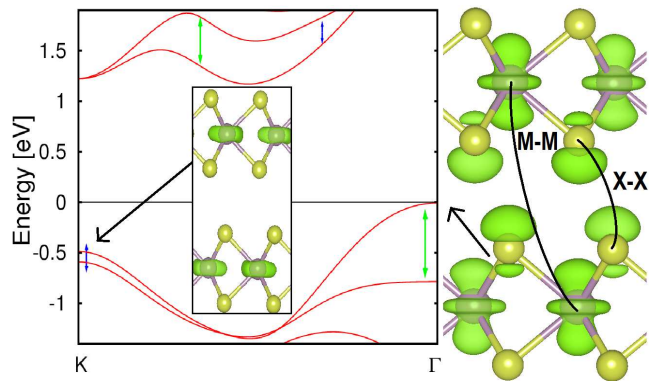}
\caption{DFT Band structure for MoS$_2$ bilayer (red line).
The green arrows represent the band splitting produced by the out of plane orbitals.
The blue arrows represent the band splitting produced by the in plane orbitals. In the inset is shown the charge density
of the last occupied band at the K point. Beside the plot we have the charge density of the last occupied band at the $\Gamma$ point.
}
\label{BandBIL}
\end{figure}
To understand the effect on the band structure, it is convenient to separate the $sym$-orbitals between the in plane orbitals (p$_x$,p$_y$,d$_{xy}$,d$_{x^2-r^2}$) and out of plane orbitals (p$_z$, d$_{3z^2-r^2}$).
For the out of plane orbitals the hybridization is larger at the $\Gamma$ point and vanishes going
towards K as in the following formula that describe the hybridization X-X'
\begin{equation}\label{BA}
  \varepsilon^{z,z'}(k_x,k_y)=t^{z,z'}\left(\mathrm{e}^{\frac{ik_ya}{\sqrt{3}}}+2\cos{\frac{k_xa}{2}}\mathrm{e}^{-\frac{ik_ya}{2\sqrt{3}}}\right)
\end{equation}
The band splitting produced by the out of plane orbitals in the V band at the $\Gamma$ point is 0.78 eV.
We can observe as the bonding-antibonding (green arrows in Fig. \ref{BandBIL}) splitting moving from the V band to the C band going from $\Gamma$ to K
thanks to the inversion of the d-bands already discussed.
The bonding-antibonding splitting caused by the out of plane orbitals is responsible for the transition from direct to indirect gap in the bilayer as we can see from the last occupied bands at the $\Gamma$ point in Fig. \ref{BandBIL}.
The same is valid for the in plane orbitals but the hybridizations behave in opposite way,
they are larger at the K point and they reduce going to the $\Gamma$ point (blue arrows in Fig. \ref{BandBIL}).
The maximum band splitting produced by the in plane orbitals is 0.10 eV in the V band at the K point.
Between the $\Gamma$ and the K points, we have a strong mixing of the in plane and out of plane orbitals,
but likely both effects should produce the splitting in the C band with predominant contribution from the
out of plane orbitals that have larger hybridizations. Indeed using just the p$_{z}$-p$_{z'}$ hybridization the splitting in the C band is underestimated\cite{Cappelluti}.
We summarize our results for the bilayer in Table \ref{TabGapBilayer}.
All these splitting effects, that reduce the gap, are expected to increase for the inner layers both in bulk and multilayers.
As these gap reducing effects are general, the gap is expected to reduce respect to the single layers for any combination of these compounds stacking faults along the $c$-axis.
\begin{table}[!ht]
\caption{The hybridizations can be favorable (F), against (A) or
almost ineffective (I) to generate the reported properties in MX$_2$ multilayers and bulk.
M$^\alpha$ represents the d-orbitals of the M atom where $\alpha$=1,2,3 are respectively x$^2$-y$^2$, xy and 3z$^2$-r$^2$.
X$^x$, X$^y$ and X$^z$ represent the p$_x$, p$_y$ and p$_z$ orbitals of the X atoms.}
\begin{center}
\renewcommand{\arraystretch}{1.7}
\begin{tabular}{|c|c|c|c|c|}
\hline
                                   &  M$^{3}$-M$^{3'}$  & M$^{1,2}$-M$^{1',2'}$ &  X$^{z}$-X$^{z'}$ &  X$^{x,y}$-X$^{x',y'}$ \\
\hline
Split V band at $\Gamma$           &    F    &  I   &  F   &   I     \\
\hline
Split V band at $K$                &    I    &  F   &  I   &   F     \\
\hline
Split C band at $\Gamma$-K         &    F    &  F   &  F   &   F     \\
\hline
\end{tabular}
\end{center}
\label{TabGapBilayer}
\end{table}

The bonding-antibonding scenario created by the M$^{xy}$-M$^{x^2-y^2}$ hybridization
is not modified for the bilayer but the adding of the hybridization between different layers
contributes to the splitting at the K point.
Though the X-X' hybridization is dominant, we can have an analytic formula for the splitting if we set it to zero.
The bilayer splitting at the K point in the limit of $t \longrightarrow \infty$ is\cite{SUPMAT4}
\begin{equation}\nonumber
  \Delta E=2\sqrt{\lambda^2+|\varepsilon^{x^2-y^2,xy'}|^2+\left|\frac{\varepsilon^{xy,xy'}-\varepsilon^{x^2-y^2,x^2-{y^2}'}}{2}\right|^2}
\end{equation}
and we can observe how the splitting has a dependence both on SOC and hybridizations.
The splitting for the bilayer is always larger than the splitting for the monolayer.
The splitting is non zero also when the SOC vanishes and it is always larger than the splitting for the monolayer.
At $\lambda$=0 using the numerical values\cite{SUPMAT4} we get $\Delta E$=23.7~meV for the MoS$_2$.
This is a small contribution compared with the spin-orbit and the
hybridization between the in plane X-p orbitals of the different monolayers. Thus, we have demonstrated that
the splitting of the valence band in the case of the bilayer does not come just from the spin-orbit coupling.

\smallskip

\section{Conclusions}

In conclusion, the model presented here provides a basis for tight-binding calculations
for MX$_2$ systems with an {\it ab-initio} accuracy.
The crystal field splitting induced by the X$^{-2}$-like atoms is barely affected while changing the X element.
On the contrary, the M-M and the M-X hybridization are reduced going down the group because of the increase in the lattice parameter.
The CF, the M-M and M-X hybridization play in the same direction to open the gap at the $\Gamma$ point.
Along the $\Gamma$-$K$ line the M$^\alpha$-M$^\alpha$ tends to create a crossing of the
$sym$-band closing the gap but the M-X hybridization opens it.
At the $K$ point, the M$^\alpha$-M$^\alpha$ hybridization inverts the $d$-bands respect to the CF and the gap is enhanced by the M-X and the M$^{xy}$-M$^{x^2-y^2}$ hybridization. We show that we need both the M-X and M-M hybridization to open the gap.
The large SO splitting observed in the V band at the $K$ point is assisted by the  strong M$^{xy}$-M$^{x^2-y^2}$ hybridization and the absence of the inversion symmetry. The change from direct to indirect gap and the splitting
in the valence bands in the multilayers is attributed
to the hybridization between out of plane orbitals of different layers. The splitting at the K point is attributed both to the SOC and to the hybridization
between in plane orbitals.

Recently, a similar paper has been published, see Ref. \onlinecite{novemberpaper}.
Though the basic computation details in the initial part of the papers are the same, the physical properties under
consideration differ in the rest of the papers.

\begin{acknowledgments}

We thank C. Noce and S. Haldar for useful discussions.
The simulations were performed on resources
provided by the Swedish National Infrastructure (SNIC) at National Supercomputer Centre at Link\"{o}ping University (NSC).

\end{acknowledgments}

\end{document}